\documentclass[aps,prl,twocolumn,groupedaddress,amssymb,amsmath,showpacs]{revtex4}
\usepackage{bm}
\usepackage{amsmath}
\usepackage{amssymb}
\usepackage{amsthm}
\usepackage{amsfonts}
\usepackage{graphicx}
\usepackage{color}
\usepackage[colorlinks=true,urlcolor=blue,linkcolor=green,citecolor=red]{hyperref}

\begin{document}


\title{\textit{Verschr\"{a}nkung} versus \textit{ Stosszahlansatz}:
Disappearance of the Thermodynamic Arrow in a High-Correlation Environment}


\author{M. Hossein Partovi}
\email[Electronic address:\,\,]{hpartovi@csus.edu}
\affiliation{Department of Physics and Astronomy, California State
University, Sacramento, California 95819-6041}


\date{\today}

\begin{abstract}
The crucial role of ambient correlations in determining thermodynamic
behavior is established. A class of entangled states of two macroscopic
systems is constructed such that each component is in a state of thermal
equilibrium at a given temperature, and when the two are allowed to interact
heat can flow from the colder to the hotter system. A dilute gas model
exhibiting this behavior is presented.  This reversal of the thermodynamic
arrow is a consequence of the entanglement between the two systems, a
condition that is opposite to molecular chaos and shown to be unlikely in a
low-entropy environment.  By contrast, the second law is established by
proving Clausius' inequality in a low-entropy environment. These general
results strongly support the expectation, first expressed by Boltzmann and
subsequently elaborated by others, that the second law is an emergent
phenomenon that requires a low-entropy cosmological environment, one that can
effectively function as an ideal information sink.

\end{abstract}


\maketitle



The status of the second law of thermodynamics and the emergence of
macroscopic irreversibility from time-symmetric dynamical laws have been
widely debated since Boltzmann's ground breaking work relating thermodynamic
behavior to microscopic dynamics late in the nineteenth century. Indeed,
rarely have so many distinguished physicists written as extensively on a
subject while achieving so little consensus.  The debate continues to present
day, having gained intensity in recent decades as a surge of activity in such
topics as chaos, quantum computing, and quantum information theory has
stimulated interest in the subject \cite{lit}.

There is nevertheless a slowly growing consensus that the asymmetry observed
in macroscopic phenomena originates in the ``initial conditions" of our
cosmic neighborhood, and ultimately that of the whole universe.  Remarkably,
Boltzmann himself arrived at the hypothesis that ``the universe, considered
as a mechanical system\textemdash or at least a very large part of it which
surrounds us\textemdash started from a very improbable state, and is still in
a very improbable state" as the initial condition that was required to
explain the asymmetry in the second law \cite{boltzmann10}. This view has
subsequently been echoed by Gold, Feynman, Lebowitz and coworkers, Peierls,
Penrose, and others \cite{luminaries}, albeit with various degrees of
emphasis and detail.  In particular, Penrose has characterized the
improbability of the initial state of the universe by the
vanishing of the Weyl tensor in the early, homogeneous universe, in contrast
to its preponderance in the final, clumped stages (which are the higher
entropy states in the presence of gravity). If so, how is the influence of a
low-entropy, time-asymmetric cosmological environment felt by a thermodynamic
system? ``This is reflected in Boltzmann's \textit{stosszahlansatz}," remarks
Peierls \cite{luminaries}, referring to Boltzmann's assumption of molecular
chaos that single particle states are uncorrelated in a dilute gas
\cite{boltzmann2}. Presumably, one would expect the initial correlations
among the molecules of the gas to be small in a low-entropy environment, thus
allowing subsequent interactions to raise their individual entropies toward
equilibrium.

Thus arises the question of whether in a high-entropy environment
interactions can lower initially high correlations among the molecules,
thereby lowering their individual entropies and reversing the thermodynamic
arrow. More specifically, can the direction of heat flow be reversed in an
environment in which entanglement is more typical than molecular chaos?  The
primary purpose of this Letter is to show that this is indeed possible.  In
particular, we construct a class of entangled states of two macroscopic
systems such that each individual component is in a state of thermal
equilibrium at a given temperature, and when the two are allowed to interact
heat can flow from the colder to the hotter system. An explicit example of
this class where the two systems are dilute gases is also constructed. The
resulting reversal of the thermodynamic arrow is thus a consequence of
entanglement between the two systems.  In effect, this entanglement disables
the statistical biases that give rise to normal thermodynamic behavior.  To
analyze the opposite scenario, we first establish the result that
correlations are generally small in a low-entropy environment, a result that
justifies the condition of generalized molecular chaos in a low-entropy
universe.  We then establish the second law by proving Clausius' inequality
in a low-entropy environment. These results strongly support the view that
the second law and the thermodynamic arrow are emergent phenomena that
require a low-entropy environment, with the universe effectively functioning
as an infinite information sink.  In our cosmic neighborhood, this is made
possible by a bright sun against a dark sky, thus maintaining a steady
process of entropy disposal \cite{various}.  Needless to say, the currently
favored accelerating models of big bang cosmology nicely accommodate the role
required of the universe.

As a preliminary step, we will establish the result that two-body correlation
is on average bounded by single-body entropy in any collection of $N \geq 3$
interacting systems. Let the systems be labeled $i=1,2,\ldots N$, and the
corresponding von Neumann entropies ${S}^{i}$.  To quantify two-body
correlations, we will use ${I}^{ij}={S}^{i}+ {S}^{j}-{S}^{ij}$, the quantum
measure of mutual information. The basic tool in the derivation is the strong
subadditivity property of entropy in the form ${S}^{i}+{S}^{j} \leq
{S}^{ik}+{S}^{jk}$ \cite{ruskai}. Since there are $N(N-1)/2$ distinct pairs
in the collection, there will be as many distinct inequalities resulting from
strong subadditivity.  When aggregated, they give
\begin{equation}
{[N(N-1)/2]}^{-1} {\sum}_{i<j=1}^{N(N-1)/2} {I}^{ij} \leq {N}^{-1}
{\sum}_{i=1}^{N} {S}^{i}. \label{1}
\end{equation}
Thus ${I}_{av} \leq {S}_{av}$, indicating that a small average entropy
guarantees a low level of two-body correlations.  The significance of this
general result is the validity of Boltzmann's \textit{stosszahlansatz}, for
any two systems whether microscopic or macroscopic, as a likely condition in
a low-entropy universe.  Note that we established this result without any
reference to such problematic issues as the purity level of the wavefunction
of the universe.

In much of what follows, we will rely on a fundamental inequality that
governs the evolution of any system whose initial state is one of thermal
equilibrium.   Let the initial and final states of the system be
${\rho}_{i}=\exp (-\beta {{H}}_{i})/Z$ and ${\rho}_{f}$ respectively, where
$\beta=1/{k}_{B}T$.  Here ${\rho}_{i}$ is the density matrix describing the
initial Gibbs state, with ${{H}}_{i}$ the initial Hamiltonian operator, $Z$
the partition function, $T$ the temperature, and ${k}_{B}$ the Boltzmann
constant.    Note that the Hamiltonian may change during the evolution,
causing exchange of work.  Thus the final Hamiltonian ${{H}}_{f}$ will in
general be different from  ${{H}}_{i}$.  The evolution itself will in general
not be unitary as it may involve interaction with other systems, and the
final state ${\rho}_{f}$ may not be one of equilibrium.

Now consider $S({\rho}_{f} \| {\rho}_{i})$, the relative entropy of
${\rho}_{f}$ with respect to ${\rho}_{i}$, which is a non-negative quantity
defined as $-S({\rho}_{f})-\textrm{tr}({\rho}_{f}\log{\rho}_{i})$.  The
latter property can be used to establish that
\begin{equation}
S({\rho}_{f} \| {\rho}_{i})=\beta \Delta U-\Delta S -  \beta \, \rm{tr}
({\rho}_{f} \Delta H) \geq 0, \label{2}
\end{equation}
where $\Delta U ={U}_{f}-{U}_{i}=\rm{tr} ({\rho}_{f} {H}_{f})-\rm{tr}
({\rho}_{i} {H}_{i})$, $\Delta S ={S}_{f}-{S}_{i}$, and
$S(\rho)=-\rm{tr}(\rho \log \rho)$ is the von Neumann entropy. It is
important to realize that this inequality stipulates an \textit{initial}
equilibrium state and temperature only, and as such is fundamentally
different from the standard inequalities of equilibrium thermodynamics.

\textit{Verschr\"{a}nkung} versus \textit{ Stosszahlansatz}.  Consider two
systems $A$ and $B$, each individually in thermal equilibrium at temperatures
${\beta}^{A}$ and ${\beta}^{B}$ initially, which are placed in thermal
contact and allowed to exchange heat but not work.  In that case $\Delta
H={H}_{f}-{H}_{i}=0$, $\Delta U$ equals the absorbed heat $Q$, and inequality
(\ref{2}) reduces to $\beta Q -\Delta S \geq 0$ \cite{PartoviQT}. Applying
the latter to each of the two systems in the above process, we find
${\beta}^{A} {Q}^{A} \geq {\Delta S}^{A} $ and ${\beta}^{B} {Q}^{B} \geq
{\Delta S}^{B} $, while ${Q}^{A}+{Q}^{B}=0$.  We will next apply these
inequalities to the extreme cases of zero and maximum initial correlations,
designating them as cases \textbf{S} and \textbf{V} respectively.

\textbf{Case S}.  If the two systems are initially uncorrelated, as is
typically expected in a low-entropy environment according to (\ref{1}), we
will have ${\rho}^{AB}_{i}={\rho}^{A}_{i} \otimes {\rho}^{B}_{i}$ and
${S}^{AB}_{i} = {S}^{A}_{i}+{S}^{B}_{i}$.  On the other hand, the final state
${\rho}^{AB}_{f}$ will generally be correlated due to the interaction and we
will have ${S}^{AB}_{f} \leq {S}^{A}_{f}+{S}^{B}_{f}$.  Since the two systems
interact in isolation from the rest of the universe, ${\rho}^{AB}_{i}$ and
${\rho}^{AB}_{f}$ will be related unitarily and we have ${S}^{AB}_{f}
={S}^{AB}_{i}$, which in conjunction with the foregoing relations leads to
${\Delta S}^{A}+{\Delta S}^{B} \geq 0$.  Combining the latter with the
inequalities of the preceding paragraph, we find ${\beta}^{A} {Q}^{A} +
{\beta}^{B} {Q}^{B} \geq 0 $. Since ${Q}^{A}+{Q}^{B}=0$, the last inequality
implies that ${Q}^{A}$ has the same sign as ${\beta}^{A}-{\beta}^{B}$ or
${T}^{B}-{T}^{A}$, i.e., that heat flows from the \textit{initially} hotter
system to the \textit{initially} colder one.  This is of course a fundamental
law of nature, and is seen here to follow where generalized
\textit{Stosszahlansatz} holds.  In essence, the foregoing argument reflects
Boltzmann's original reasoning: starting from an uncorrelated initial
configuration, there are overwhelmingly more possibilities for final states
if the initially colder state gains energy than vice versa. Model
calculations using molecules as interacting thermodynamic systems verify this
expectation in detail \cite{PartoviQT}.

\textbf{Case V}.  The above scenario changes dramatically if the two
interacting systems are significantly correlated to begin with, as may be
expected in a high-correlation environment \cite{ME}.  To demonstrate this
assertion, we will consider the extreme case where the two systems are
initially entangled in a pure state while each individual system is in
thermal equilibrium.  These conditions can be realized for a pair of
macroscopic systems whose energy spectra, $\{ {E}_{i}^{A} \}$ and $\{
{E}_{i}^{B} \}$, are identical except for a scale factor, i.e., if ${\mu}^{A}
{E}_{i}^{A}={\mu}^{B} {E}_{i}^{B} = {\epsilon}_{i} $.  The desired joint
state of the two systems can then be represented as
${\rho}^{AB}=|{\Omega}^{AB} \rangle \langle {\Omega}^{AB}|$, with
\begin{equation}
| {\Omega}^{AB}\rangle={Z}^{-1/2} {\sum}_{i} \exp (- \gamma
{\epsilon}_{i}/2)|i;A \rangle |i;B \rangle , \label{3}
\end{equation}
where $|i;A \rangle$ ($|i;B \rangle$) is the $i$th energy eigenvector for
system $A$ ($B$), $\gamma$ is a positive constant, and ${Z}^{-1/2}$ is a
normalization constant.  Note that Eq.~(\ref{3}) is essentially the Schmidt
decomposition of $|{\Omega}^{AB}\rangle$.  It can now be readily verified
that the individual states of the two systems (obtained from ${\rho}^{AB}$ by
tracing over the Hilbert space of the other) are thermal equilibrium states
at temperatures given by ${\beta}^{A}={\mu}^{A} \gamma$ and
${\beta}^{B}={\mu}^{B} \gamma$.

Just as in case \textbf{S}, we consider a process of heat exchange between
$A$ and $B$ and find ${\beta}^{A}{Q}^{A} \geq {\Delta S}^{A}$,
${\beta}^{B}{Q}^{B} \geq {\Delta S}^{B}$, and ${Q}^{A}+{Q}^{B}=0$.  However,
in contrast to case \textbf{S}, the joint state of the two systems is pure in
this case, so that ${\rho}^{A}$ and ${\rho}^{B}$ are now isospectral with the
consequence that ${S}^{A}={ S}^{B}$ at all times and ${\Delta S}^{A} ={\Delta
S}^{B}$ for the process.  Since in general ${Q}^{A}{Q}^{B}\leq 0$, the above
inequalities imply that ${\Delta S}^{A}={\Delta S}^{B} \leq 0$, thus
reversing the inequality we found for ${\Delta S}^{A}+{\Delta S}^{B}$ in case
\textbf{S}.  This reversal in turn leads to ${\beta}^{A} {Q}^{A} +
{\beta}^{B} {Q}^{B} \geq {\Delta S}^{A}+{\Delta S}^{B}$, which allows both
directions of heat flow, including that from the initially colder body to the
hotter one.  Note that equality in this result obtains only if there is zero
heat exchange between the systems (in violation of the zeroth law since
initial temperatures are unrestricted here).  We will later present a model
exhibiting this reversal explicitly.

What is the cause of this bizarre behavior?  The clue is in the dual
character of $|{\Omega}^{AB}\rangle$: while entanglement forces the
individual entropies ${ S}^{A}$ and ${ S}^{B}$ to move in lock-step, the
maximal entropy of the initial equilibrium states implies that the individual
entropies can only decrease.  Therefore, there is no opportunity for any
statistical dominance of one direction of heat flow over the other, in stark
contrast to case \textbf{S}.  This feature is strikingly apparent in the
model calculation considered later.  The main lesson to be drawn here is that
the statistical biases that cause normal thermodynamic behavior can be
neutralized by pre-existing correlations between the interacting systems.
Thus a low-entropy environment, which serves to guarantee low correlations
and preclude the anomalous behavior just described, is indeed a pre-requisite
for normal thermodynamic behavior.

To further highlight the contrast between the two cases considered above, we
will now establish the second law in case \textbf{S} by proving Clausius'
Theorem in a low-entropy environment.  Consider a macroscopic system that
undergoes a cyclic evolution in thermal contact with a series of heat
reservoirs, absorbing ${Q}_{j}^{S}$ from the reservoir at temperature
${T}^{R}_{j}$ while exchanging work as a result of possible changes in its
Hamiltonian (e.g., because of expansion).  Note that by definition the work
exchange is associated with a unitary evolution and does not entail
information transfer, in contrast to the heat exchange with the reservoir.
Note also that ${T}_{j}$ is the temperature of the $j$th reservoir, as
there's no presumption of thermal equilibrium with the system here.

Consider the $j$th process, starting with the uncorrelated system-reservoir
state ${\rho}^{S}_{j} {\rho}^{R}_{j}$, where ${\rho}^{R}_{j}$ is a Gibbs
state at temperature ${T}_{j}^{R}$ and ${\rho}^{S}_{j}$ is arbitrary, and
culminating in the correlated state ${\rho}^{SR}_{j}$.  The Hamiltonian
operator governing this evolution may be represented as
${H}^{S}_{j}+{V}^{S}_{j}+{H}^{R}_{j}+{H}^{SR}_{j}$, where ${H}^{S}_{j}$ and
${H}^{R}_{j}$ refer to the system and reservoir, ${H}^{SR}_{j}$ to their
interaction, and ${V}^{S}_{j}$ to the interaction of the system with the
external agents with which it is exchanging work.  Just as in case \textbf{S}
above, we find ${\Delta S}_{j}^{S}+{\Delta S}^{R}_{j} \geq 0$, ${V}^{S}_{j}$
notwithstanding, and applying inequality (\ref{2}) to the \textit{reservoir},
we conclude that ${\beta}^{R}_{j} {Q}^{R}_{j} \geq {\Delta S}^{R}_{j} $,
where ${\beta}^{R}_{j}=1/k{T}^{R}_{j}$ refers to the initial temperature of
the reservoir.  Furthermore, the energy exchange between the system and the
reservoir is subject to  ${Q}_{j}^{R}=-{Q}_{j}^{S}$, since the work exchange
does not involve the reservoir.  Combining the last two inequalities, we find
${\beta}_{j}^{R} {Q}_{j}^{S} + {\Delta S}_{j}^{S} \leq 0$.  Remarkably, the
system obeys this inequality without necessarily having a well-defined
temperature (${\beta}_{j}^{R}$ refers to the reservoir) and regardless of
possible changes in its Hamiltonian.

If the last inequality is summed over the cycle, the entropy changes add up
to zero, since the final state of the system is the same as the initial one,
and we find
\begin{equation}
{\sum}_{j} {\beta}_{j} {Q}_{j}^{S} \leq 0, \label{4}
\end{equation}
which is Clausius' inequality.

Next we will construct a dilute-gas model for case \textbf{V} that will
exhibit heat flow from the colder to the hotter system.  The dilute nature of
the two gases allows us to simplify the calculation by focusing attention on
single particle interactions.  We therefore consider particle $a$ of one gas
interacting with particle $b$ of the other in a volume $V$.  The initial,
joint state of the two is pure and entangled, but in such a way that their
individual states are in thermal equilibrium at different temperatures.  To
exhibit the structure of this state, we let $|a,{\alpha}_{a}
\mathbf{k}\rangle$ represent a state of momentum ${\alpha}_{a} \mathbf{k}$
for particle $a$ and $|b,{\alpha}_{b} \mathbf{k}\rangle$ a state of momentum
${\alpha}_{b} \mathbf{k}$ for particle $b$, where ${\alpha}_{a}$ and
${\alpha}_{b}$ are positive parameters.  The joint state of the two particles
is then
\begin{equation}
|{\omega}^{ab}\rangle={Z}^{-1/2} {\sum}_{\mathbf{k}} \exp (- \gamma
{\mathbf{k}}^{2}/4m) |a,{\alpha}_{a} \mathbf{k}\rangle |b,{\alpha}_{b}
\mathbf{k}\rangle, \label{5}
\end{equation}
where $Z$ and $\gamma$ are as defined in Eq.~(\ref{3}), $m$ is a mass scale,
and we have set $\hbar=1$.

It is useful at this point to continue the calculation using the
configuration representation in the infinite-volume limit.  Then
Eq.~(\ref{5}) appears as
\begin{eqnarray}
{\omega}^{ab}({\mathbf{r}}^{a},{\mathbf{r}}^{b}) & = {[\sqrt{Z} {(2 \pi
)}^{3}]}^{-1} \int {d}^{3}k  \exp (- \gamma {\mathbf{k}}^{2}/4m) \nonumber \\
&\times \exp (i{\alpha}_{a} \mathbf{k}\cdot {\mathbf{r}}^{a})  \exp
(i{\alpha}_{b} \mathbf{k}\cdot {\mathbf{r}}^{b}). \label{6}
\end{eqnarray}
The individual (or marginal) states of the two systems can now be found as
\begin{eqnarray}
{\rho}^{a}(\mathbf{r},\mathbf{r'})&={[{Z} {(2 \pi {\alpha}_{a}
{\alpha}_{b})}^{3}]}^{-1} \int {d}^{3}k
\exp (- \gamma {\mathbf{k}}^{2}/2m{\alpha}_{a}^{2}) \nonumber \\
& \times \exp[ i \mathbf{k}\cdot ({\mathbf{r}}-{\mathbf{r'}}) ], \label{7}
\end{eqnarray}
and an analogous expression for ${\rho}^{b}$.  Note that $\gamma
{\mathbf{k}}^{2}/2m {\alpha}_{a}^{2}={\beta}^{a}  {\mathbf{k}}^{2}/2{m}^{a}$,
which identifies ${\rho}^{a}$ as a Gibbs state at temperature
${T}^{a}=({k}_{B}{\beta}^{a})^{-1}= m{\alpha}_{a}^{2}/\gamma {k}_{B}
{m}^{a}$, and similarly for ${\rho}^{b}$ with ${T}^{b}=
m{\alpha}_{b}^{2}/\gamma{k}_{B}  {m}^{b}$, where ${m}^{a}$ and ${m}^{b}$ are
the respective masses. Thus each system, if experimented upon in isolation
from the other, will be found to be in thermal equilibrium at the specified
temperature. However, if the two systems are allowed to interact, it is the
pure state given in Eq.~(\ref{6}) that must be considered as their initial
state.

To model a thermal interaction between the two systems, we consider an
adiabatic switching of the interaction so that the initial and final states
are non-interacting and the interaction of the two with external systems is
made negligible.  Such an interaction amounts to a collision, with
${\omega}^{ab}({\mathbf{r}}^{a},{\mathbf{r}}^{b})$ representing the
non-interacting incoming state.  Therefore, the fully interacting state of
the two-body system can be represented as
\begin{eqnarray}
{\omega}^{ab}_{int}({\mathbf{r}}^{a},{\mathbf{r}}^{b}) & = {[\sqrt{Z} {(2 \pi
)}^{3}]}^{-1} \int {d}^{3}k \exp (- \gamma {\mathbf{k}}^{2}/4m) \nonumber \\
&\times {\psi}_{\mathbf{k}}^{+}({\mathbf{r}}^{a},{\mathbf{r}}^{b}), \label{8}
\end{eqnarray}
where ${\psi}_{\mathbf{k}}^{+}({\mathbf{r}}^{a},{\mathbf{r}}^{b})$ is the
``in'' state that corresponds to the non-interacting state $\exp
(i{\alpha}_{a} \mathbf{k}\cdot {\mathbf{r}}^{a})  \exp(i{\alpha}_{b}
\mathbf{k}\cdot {\mathbf{r}}^{b})$.  Note that the initial momenta of the
particles in each term of the coherent superposition
${\omega}^{ab}_{int}({\mathbf{r}}^{a},{\mathbf{r}}^{b})$ are collinear and
proportional in magnitude, reflecting the entangled nature of this state.  It
is important to realize that there would be no such correlation between the
initial momenta of the particles otherwise. For example, the uncorrelated
initial state ${\rho}^{a}\otimes{\rho}^{b}$, which would be typical in a
low-entropy environment and exhibit normal thermodynamic behavior as shown
above for case \textbf{S}, entails particle pairs of uncorrelated momenta.

The correlation between the initial momenta of the colliding particles in
${\omega}^{ab}_{int}({\mathbf{r}}^{a},{\mathbf{r}}^{b})$ makes it possible to
determine the direction of energy flow from conservation laws. For a
collision event that starts with momenta ${\alpha}_{a} \mathbf{k}$ and
${\alpha}_{b} \mathbf{k}$, we find that the fractional kinetic energy gain
for particle $a$ as a result of the collision is given by
$4x(x-1){\sin}^{2}(\theta/2)$, where
$x=[{m}^{a}/({m}^{a}+{m}^{b})][({\alpha}_{a}+{\alpha}_{b})/{\alpha}_{a}]$,
and $\theta$ is the scattering angle in the center-of-mass system.  Thus for
all but forward scattering, particle $a$ gains energy if $x>1$.  This
condition can be satisfied if  ${({m}^{a}/{m}^{b})({T}^{b}/{T}^{a})}>1$,
where we have eliminated ${\alpha}_{a,b}$ in favor of the initial
temperatures.   Clearly, if ${{m}^{a}}>{{m}^{b}}$, particle $a$ can gain
energy even if ${T}^{a} >{T}^{b}$, thus reversing the normal direction of
heat flow.  Note that the direction of energy flow would then be the same for
every term in the wave packet of Eq.~(\ref{8}), clearly indicating that
thermal statistics play no role in this result.  This is the main lesson of
this example, to wit, that entanglement is capable of destroying normal
thermodynamic behavior by defeating the statistical biases that underlie it,
instead rendering such macroscopic outcomes as the direction of heat flow
dependent on microscopic details.

We close this Letter with a few concluding remarks.  First, we have focused
on the two extremes of ambient correlations, cases \textbf{S} and \textbf{V},
primarily to underscore the contrast in their resulting thermodynamic
behavior. To be sure, these extreme cases are idealizations.  However, they
do serve to characterize likely thermodynamic behavior under the specified
conditions of ambient correlations.  Second, while we have used the von
Neumann entropy extensively in our calculations,  the main results of our
analysis concern energy flow and are not committed to any specific
interpretation of that quantity for non-equilibrium states. Rather, the
primary role of the entropy function in our analysis is as a measure of
information, which is inextricably intertwined with energy flow in
thermodynamic interactions. Third, it may be noted that the results presented
here apply to the thermodynamics of microscopic systems as well. However, the
two regimes are distinguished by the magnitude of fluctuations, which are
comparable to averages for microscopic systems and normally negligible for
macroscopic systems.  Finally, it is important to note the importance of
using the quantum description where entanglement effects play an essential
role. Thus we have avoided classical phase space methods as inadvisable in
the present context, although it may be possible to formulate some of the
arguments presented here in classical terms.

{}

\end{document}